# Increases in the Irreversibility Field and the Upper Critical Field of Bulk MgB$_2$ by ZrB$_2$ Addition

M. Bhatia[1], M.D.Sumption[1], E.W. Collings[1], S.Dregia[2]

[1]LASM, Materials Science and Engineering Department,

[2] Materials Science and Engineering Department

OSU, Columbus, OH 43210, USA

*Abstract*

   In a study of the influence of ZrB$_2$ additions on the irreversibility field, $\mu_oH_{irr}$ and the upper critical field $B_{c2}$, bulk samples with 7.5 at. % ZrB$_2$ additions were made by a powder milling and compaction technique. These samples were then heated to 700-900$^o$C for 0.5 hours. Resistive transitions were measured at 4.2 K and $\mu_oH_{irr}$ and $B_{c2}$ values were determined. An increase in $B_{c2}$ from 20.5 T to 28.6 T and enhancement of $\mu_oH_{irr}$ from 16 T to 24 T were observed in the ZrB$_2$ doped sample as compared to the binary sample at 4.2 K. Critical field increases similar to those found with SiC doping were seen at 4.2 K. At higher temperatures, increases in $\mu_oH_{irr}$ were also determined by *M-H* loop extrapolation and closure. Values of $\mu_oH_{irr}$ which were enhanced with ZrB$_2$ doping (as compared to the binary) were seen at temperatures up to 34 K, with $\mu_oH_{irr}$ values larger than those for SiC doped samples at higher temperatures. The transition temperature, $T_c$, was then measured using DC susceptibility and a 2.5 K drop of the midpoint of $T_c$ was observed. The critical current density was determined using magnetic measurements and was found to increase at all temperatures between 4.2 K and 35 K with ZrB$_2$ doping.

*Keywords*: MgB$_2$, ZrB$_2$ doping,  $H_{irr}$, $B_{c2}$

Magnesium diboride, a simple binary compound, has attracted considerable attention from the scientific community in terms of fundamental and applied research. Various attempts have been made to raise its irreversibility field, $\mu_0H_{irr}$, upper critical field, $B_{c2}$, and critical current density $J_c$. It has been demonstrated that in the case of dirty $MgB_2$ thin films, $\mu_0H_{irr}$ and $B_{c2}$ can be markedly enhanced as compared to the pure binary films. This has been most evident in C doped thin film results[1,2] with $B_{c2}$ reaching 49 T at 4.2 K. This effect is understood to be due to possible lattice distortions caused by the C substitution in the B-sublattice[2]. These lattice distortions may lead to increased phonon scattering. As compared to other superconductors, this effect is more pronounced in case of $MgB_2$ because of its two-gap nature[3]. A full understanding of the material modifications that enable the $B_{c2}$ enhancement in $MgB_2$ is still lacking but a number of researchers have investigated C doping in bulk $MgB_2$ in an effort to generate the results similar to $B_{c2}$ enhanced thin films[4-5].

Apart from C doping various other dopants and preparation methods have been investigated leading to wide variation in the reported values of $\mu_0H_{irr}$ and $B_{c2}$ based on the preparation conditions. Irreversibility fields for *ex-situ* $MgB_2$ tapes have been reported to be 12 T for field perpendicular to the tape face[6] and around 16 T for the parallel fields. Marked enhancement of $\mu_0H_{irr}$ and high field $J_c$ in $MgB_2$ were achieved by proton irradiation introduced atomic disorder[7].

Dou *et al.*, in a series of papers[8-11], have shown that SiC and C nanoparticle doping can significantly improve $\mu_0H_{irr}$ and $J_c$. This effect has been attributed to enhanced flux pinning induced by a combination of substitution-induced defects and highly dispersed additives. SiC doped $MgB_2$ has also been used by Sumption et al[12] in metal sheathed strands reaching $B_{c2}$ values up to 33 T. In that work, strands of similar construction were investigated systematically, and high field resistive transitions were used to demonstrate relatively large values of $\mu_0H_{irr}$ and $B_{c2}$ with various kinds of SiC dopants under various heating conditions. Matsumoto *et. al.*[13] used $SiO_2$ and SiC in the *in-situ* process, enhancing $\mu_0H_{irr}$ from $\approx 17$ T to $\approx 23$ T at 4 K. Recently Ma *et. al* [14] added 5% of the dopants $ZrSi_2$, $ZrB_2$ and $WSi_2$. They focused on their influence on $J_c$, attributing the observed increases to improved densification and intergrain connectivity.

However, it is useful to separate those improvements in $J_c$ stemming from increases in $\mu_0 H_{irr}$ and $B_{c2}$ and those stemming from increased pinning. Thus, to further explore the effect of transition metal diboride additions we have added $ZrB_2$ into the $MgB_2$ and performed magnetic and transport measurements on bulk pellets, aimed at determining $\mu_0 H_{irr}$ and $B_{c2}$. The results of $ZrB_2$ doping are reported in this paper and compared to the effects of SiC.

Bulk samples with compositions $MgB_2$ and $(MgB_2)_{0.925}(ZrB_2)_{0.075}$ were prepared by an in-situ reaction of a stoichiometric mixture of 99.9 % pure Mg and amorphous B powders with a typical size of 1-2 μm. Powders were mixed in SPEX mill for 48 mins. Doping was achieved by adding 7.5 mol% of $ZrB_2$ (from Alfa Aesar) prior to the mixing. Powders with SiC were doping were also fabricated for comparison. The milled powder was then compacted in the form of a cylindrical pellet in a steel die. These pellets were heat-treated in a steel holder, encapsulated in a quartz tube under 200 torr of Ar. Details of the sample composition, and heat-treatment are given in Table I. After sintering at temperatures of from 700-900 deg C for 0.5 hours, the cylindrical pellets were reshaped into cuboids for property measurements.

Magnetization measurements were performed from 4.2 K to 40 K on these samples using a vibrating sample magnetometer (VSM) with field sweep amplitude of 1.7 T and a sweep rate of 0.07 T/s. Susceptibility, $\chi_{dc,}$ vs. temperature, $T$ measurements were performed using a 50 mT field sweep amplitude after initial zero field cooling. Magnetically determined critical current density, $J_{c,}$ at various temperatures was extracted from the magnetization loop using the Bean model[15].

At 4.2 K, $\mu_0 H_{irr}$, and $B_{c2}$ were determined by resistive transitions with applied field, with the measurements being performed at the National High Magnetic Field Laboratory (NHMFL) Tallahassee. A four point resistance measurement was performed, using silver paste to connect the leads, and the distance between the voltage taps was 5 mm. The applied current was 10 mA, and current reversal was used. All measurements were made at 4.2 K in applied fields ranging from 0 to 33 T. The samples were placed perpendicular to the applied field, values of $\mu_0 H_{irr}$ and $B_{c2}$ being obtained taking the 10% and 90% points of the resistive transition.

At higher temperatures (20 K and 30 K) $\mu_0 H_{irr}$ was calculated two different ways. At higher temperatures where $M$-$H$ loop closure could be observed, the loop closure itself defined $\mu_0 H_{irr}$. When loop closure could not be directly observed (e.g., at somewhat lower temperatures) the quantity $\Delta M^{1/2} B^{1/4}$ was plotted vs $B$ and extrapolated to the horizontal axis to obtain an estimate of $\mu_0 H_{irr}$. This technique (a Kramer extrapolation) is usually a very good estimate of where transport current vanishes (i.e., $\mu_0 H_{irr}$) for intergrain- and grain boundary pinned conductors, although significant inhomogeneities may cause concave or convex curvature at very low currents[16.] Nevertheless, the numbers so extracted are representative of the majority of current paths in the material. XRD measurements were taken with a Sintac XDS-500 instrument using Cu K-α radiation between the $2\theta$ values of 30-70 degrees.

We have earlier studied binary $MgB_2$ bulk samples and have reported the optimized heating schedule for such samples[17]. In the present experiment the previously reported 700°C/30 min heat-treatment was used for the binary samples and the $ZrB_2$ doped samples were heated at temperatures of 700°C, 800°C and 900°C for 30 minutes. Figure 1 shows the plot of DC susceptibility vs $T$ for these samples in comparison with the binary sample. It can be seen that the $T_c$ drops by 2.5 K for sample MBZr700 as compared to MB700. Comparing this to the results of Ma et al[14] we note that while they see a drop of only 1 K, their binary sample has a lower $T_c$ than the present binary. Thus, the midpoint $T_c$ values for our $ZrB_2$ samples are very similar to that of Ma et al.

Magnetic critical current density, $J_c$, for the pure sample and the $ZrB_2$ doped samples was also measured at temperatures ranging from 10 K to 40 K; the 20 and 30 K results are shown in Figure 2. The $J_c$ of MBZr700 is a factor of two higher than the pure sample MB700 at both 20 K and 30 K. A similar trend was found for all the temperatures, which is consistent with the increases seen in the 2-12 K range by Ma et al[14].

Resistance vs. applied magnetic field curves for the binary and the $ZrB_2$ doped $MgB_2$ samples heated to 700°C for 30 minutes are shown in Figure 3 and also Table I which summarizes the heating schedules and compositions of the samples along with the measured critical fields. It can be seen that the $B_{c2}$ of the doped sample is found to be 28.6 T as compared to 20.5 T for the binary sample. This increase in $B_{c2}$ is comparable to the high $B_{c2}$ values achieved by Sumption *et al.*[12] with SiC doping. Additionally, $\mu_o H_{irr}$ has also been found to increase from 16 T to 24 T at 4.2 K with $ZrB_2$ doping. This increase in the critical fields is believed to be due to the small substitution of Zr on the Mg sites as evidenced by the change in lattice parameter, Figure 4, the effects of which are believed to be changes in the electron diffusivities in the $\pi$ and $\sigma$ band[3]. As can be seen from the XRD patterns, Figure 4, of the control sample and MBZr700 the shift of the latter's [002] and [110] peaks to lower angles suggests an increases in both *a* and *c* lattice parameters indicative of uniform lattice strain induced by Zr substitution on the Mg site. Microscopic studies to verify the substitution are under way.

A number of dopants have been seen to be effective in increasing the critical fields of $MgB_2$ at 4.2 K. One of the important questions for $MgB_2$ development is how these various dopants will perform at higher temperatures. Dopants which generate σ-band scattering are expected to perform better at higher temperatures, although it is not entirely clear which substitutions will do this. In our case we looked to characterize the critical field enhancement at a range of temperatures for $ZrB_2$ additions. At higher temperatures, direct *M-H* loop closure was used to determine $\mu_o H_{irr}$, the results are shown in Figure 5. At intermediate *T*, the Kramer extrapolation described before was used to determine $\mu_o H_{irr}$, these are also shown in Figure 5, where the 4.2 K resistance measurements are also included. An increase in the $\mu_o H_{irr}$ with $ZrB_2$ doping of $MgB_2$ was observed over the entire temperature range, with $ZrB_2$ doping enhancements more significant that those of SiC at higher temperatures.

Additions of 7.5 mol% of $ZrB_2$ into $MgB_2$ enhances $\mu_o H_{irr}$, $B_{c2}$ and $J_c$ in pellet samples at all temperatures. Resistive transitions in field at 4.2 K found a pronounced increase in $B_{c2}$ from 20.5 T for pure sample to 28.6 T for the doped sample and enhancement of $\mu_o H_{irr}$ from 16 T to 24 T. At higher temperatures magnetization loop extrapolations and direct loop closures shown $\mu_o H_{irr}$ increases as well. Values of $\mu_o H_{irr}$ which were enhanced with $ZrB_2$ doping (as compared to the binary) were seen at temperatures up to 34 K, with $\mu_o H_{irr}$ values larger than those for SiC doped samples at higher temperatures. A $T_c$ midpoint depression of 2.5 K was seen as compared to the binary sample. A two-fold increase in magnetic $J_c$ was found over the temperature range from 10-30 K in the $ZrB_2$ doped $MgB_2$.

**Acknowledgements**

This work was supported by The State of Ohio Technology Action Fund Grant and by the US Department of Energy, HEP, Grant No. DE-FG02-95ER40900. A portion of this work was performed at the National High Magnetic Field Laboratory, which is supported by NSF Cooperative Agreement No. DMR-0084173 and by the State of Florida.

**List of Tables**

Table I. Sample composition and properties.

| Sample Name | Composition | HT ($^O$C/Min) | $T_c$, K | $J_c$(20K/1T) ($10^5$A/cm$^2$) | $\mu_0H_{irr}$ T | $B_{c2}$ T |
|---|---|---|---|---|---|---|
| MB700 | Pure MgB$_2$ | 700/30 | 38.2 | 0.7 | 16.0 | 20.5 |
| MBZr700 | (MgB$_2$)$_{0.925}$(ZrB$_2$)$_{0.075}$ | 700/30 | 35.7 | 2.0 | 24.0 | 28.6 |
| MBZr800 | (MgB$_2$)$_{0.925}$(ZrB$_2$)$_{0.075}$ | 800/30 | 36.5 | 1.1 | - | - |
| MBZr900 | (MgB$_2$)$_{0.925}$(ZrB$_2$)$_{0.075}$ | 900/30 | 36.5 | 1.0 | - | - |
| MBSiC700 | (MgB$_2$)$_{0.90}$(SiC)$_{0.10}$ | 700/30 | 35.3 | 0.5 | 24.5 | >33 |

**List of Figures**

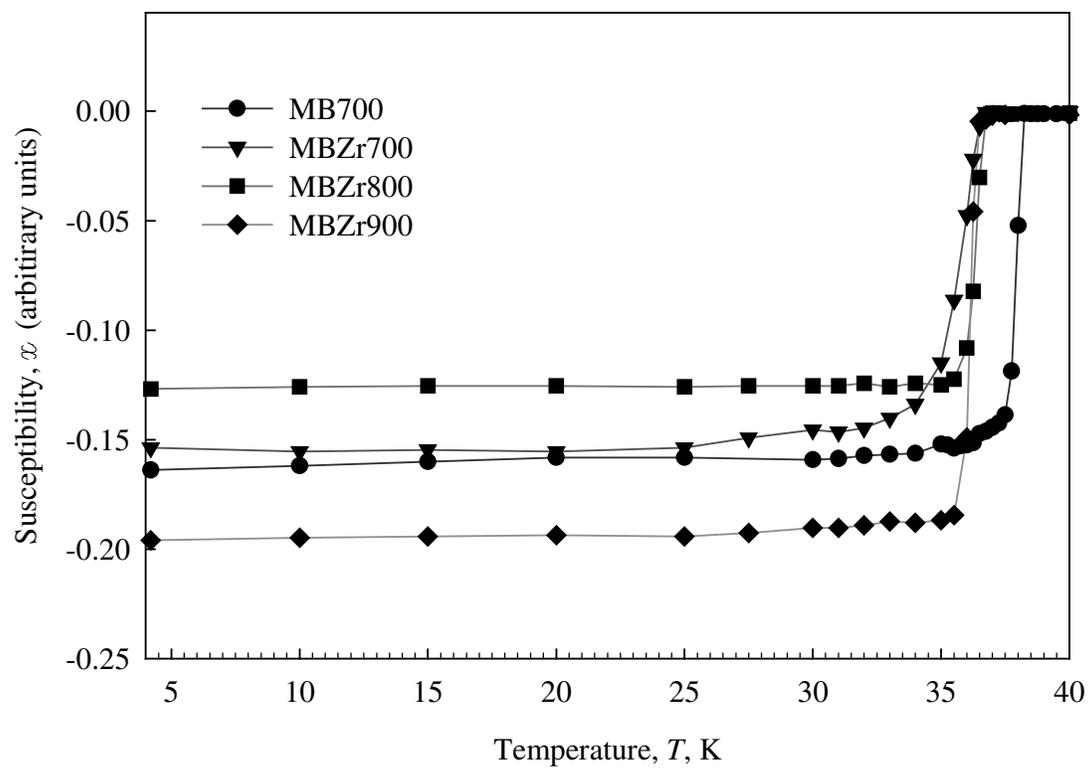

**Figure 1. Bhatia**

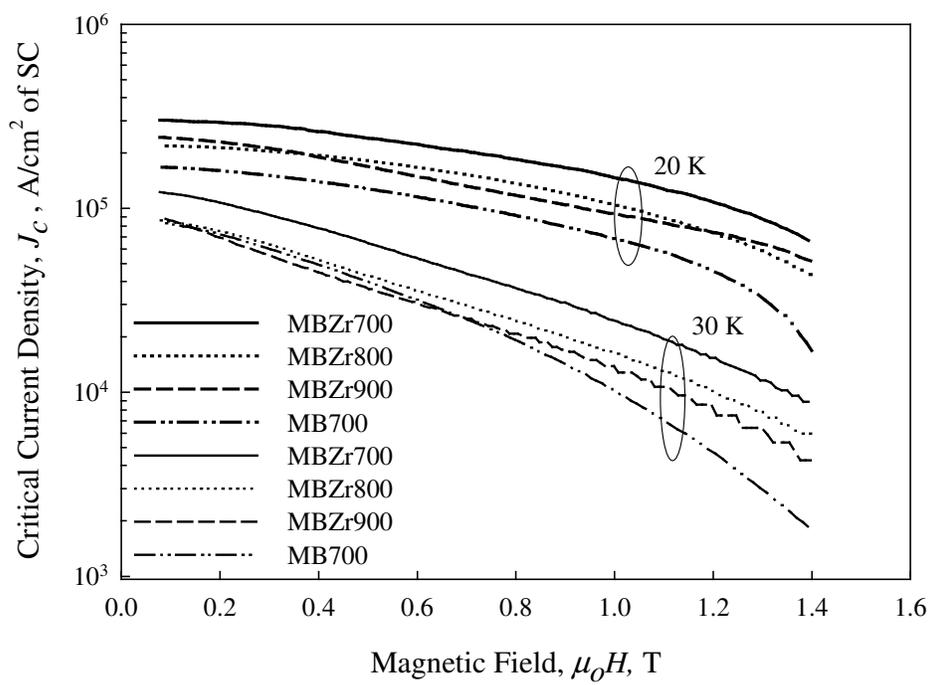

**Figure 2. Bhatia**

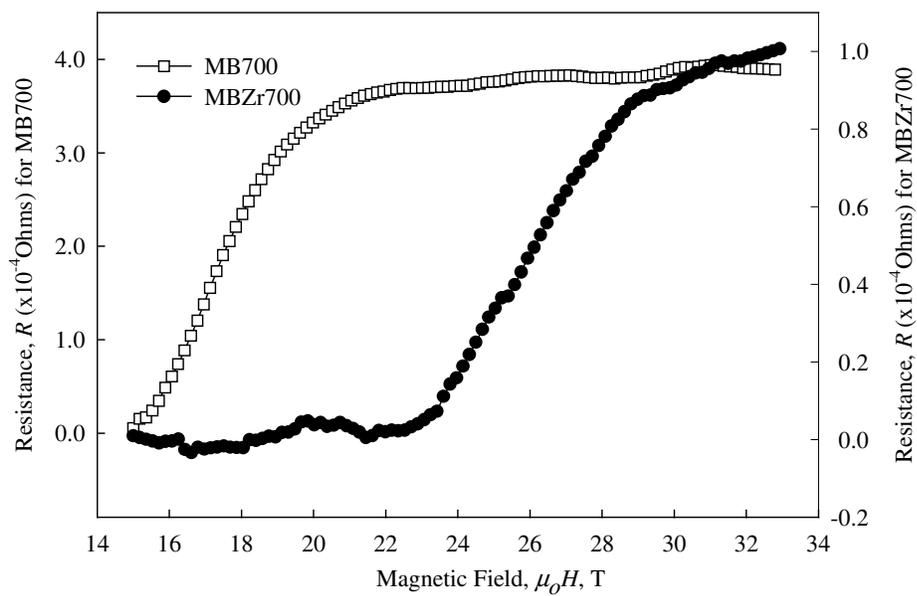

**Figure 3. Bhatia**

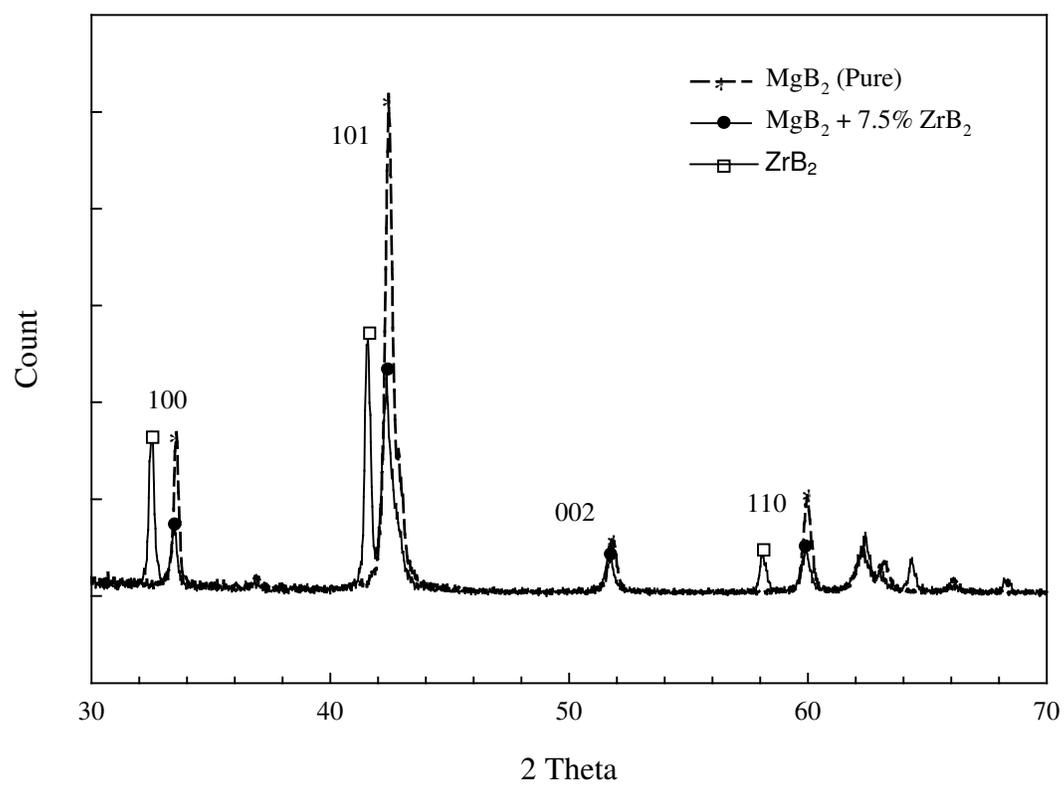

**Figure 4. Bhatia**

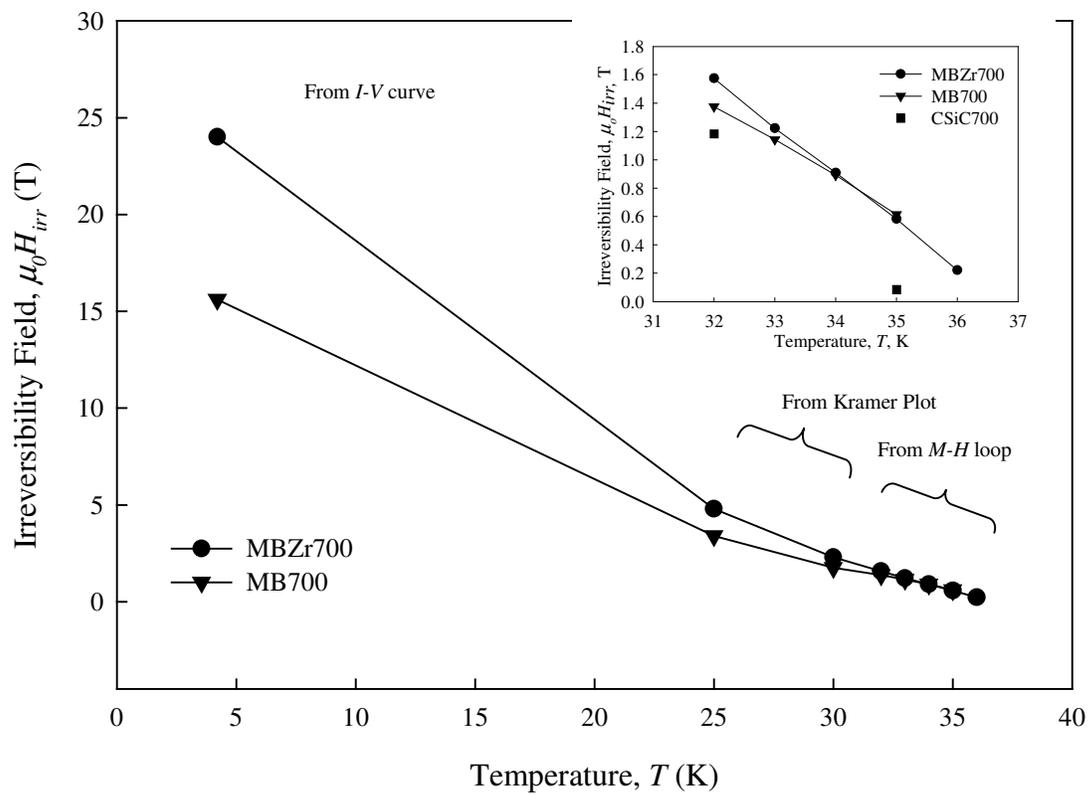

**Figure 5. Bhatia**